\documentclass[aps,prb,twocolumn,longbibliography]{revtex4-1}
\usepackage{epsfig,amsopn}
\usepackage{graphicx}
\usepackage{color}
\usepackage{amsmath,amssymb}
\usepackage{enumerate}
\newcommand\bea{\begin{eqnarray}}
\newcommand\eea{\end{eqnarray}}
\newcommand\beq{\begin{equation}}
\newcommand\eeq{\end{equation}}

\def\nn{\nonumber}
\def\f{\frac}

\def\si{\sigma}
\def\Do{\partial}
\def\De{\Delta}

\def\ua{\uparrow}
\def\da{\downarrow}

\begin{document}
\title{Fabry-P\'erot interference in Josephson junctions}
\author{ Sushil Kumar Sahu}
\author{ Abhiram Soori~~}  
\email{abhirams@uohyd.ac.in}
\affiliation{ School of Physics, University of Hyderabad, Prof. C. R. Rao Road, Gachibowli, Hyderabad-500046, India.}
\begin{abstract}
Conductance of metallic heterostructures can be controlled by applying a gate voltage to a region in the transport channel. For sufficiently long phase coherent channels, oscillations appear in conductance versus chemical potential plot, which can be explained by  Fabry-P\'erot interference. In this work, we study DC Josephson effect in a superconductor-normal metal-superconductor junctions. The chemical potential of the normal metal (NM) region can be tuned by an applied gate voltage. We numerically obtain the Andreev bound states formed within the superconducting gap and calculate Josephson current by summing up the currents carried by the occupied Andreev bound states. We find that the Josephson current oscillates as a function of the chemical potential in the NM region, and these oscillations can be explained by Fabry-P\'erot interference condition. We find that Josephson current carried by one bound state can be higher than that carried by two or more bound states. 
\end{abstract}
%\pacs{}
\maketitle

\section{Introduction}
Fabry-P\'erot interference (FPI) is a phenomenon in light scattering that happens in optical cavities wherein for certain sizes of the cavity the transmission of monochromatic light is perfect and the light is reflected from the cavity otherwise~\cite{perot}. This phenomenon has been used  to assist lasing action in lasers~\cite{Williams2007} and  in gravitational wave detectors~\cite{Kawamura94,Mik19}. The same phenomenon is exhibited by electrons in nanostructures owing to their wave nature~\cite{liang2001}. The physics of FPI is used in the detection of fractional charges in quantum Hall devices~\cite{ofek2010}. Spin transistors~\cite{soori12,sahoo2023} and  planar Hall effect devices~\cite{suri21,soori2021} exhibit FPI. Several proposals to enhance crossed Andreev reflection make use of FPI~\cite{soori17,nehra19,soori19,soori22car}. Scattering across PT-symmetric non-Hermitian ladders exhibits FPI in the PT-unbroken phase whereas FPI is absent in PT-broken phase~\cite{soori2022nh}. 

DC Josephson effect is an equilibrium phenomenon in junctions between two superconductors, wherein a current flows from one superconductor to the other when a superconducting  phase difference is maintained between  two superconductors~\cite{josephson62}. This current termed as Josephson current is carried by Cooper pairs and flows even when  a normal metal or a thin insulating layer is sandwiched  between the two superconductors. In presence of a phase difference, quasiparticle bound states appear within the superconducting gap and unlike the bound states in  normal metal which carry no current, these bound states carry the supercurrent~\cite{furusaki99}. It may be noted here that bound states can also appear in absence of a phase difference if a sufficiently long normal metal region is inserted between the superconductors, but they do not carry any supercurrent in absence of a phase bias. Josephson current through a quantum point contact has been studied by Beenakker and van Houten wherein the dependence of the Josephson current on the width of the point contact is investigated~\cite{beenakker91,vanhouten91}. In a superconductor-normal~metal-superconductor junction, the transport is coherent only when the size of the normal metal is smaller than the  coherence length. 

\begin{figure}[htb]
\includegraphics[width=9cm]{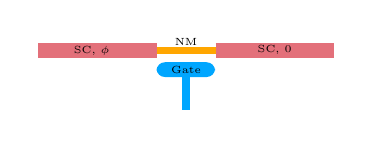}
\caption{Schematic of the setup. The superconductor (SC) on the left (right) has a phase $\phi$ ($0$). Gate voltage applied to the central normal metal (NM) can change its chemical potential. }\label{fig:schem}
\end{figure}

In this work, we study the effect of a gate tunable normal metal sandwiched between two superconductors on the Josephson current. The schematic of the setup is shown in Fig.~\ref{fig:schem}. When the chemical potential of a normal metal coupled to normal metal reservoirs on either side is changed, the differential conductance of the setup exhibits oscillations which are rooted in Fabry-P\'erot interference. Oscillations are expected in Josephson current as well when the chemical potential of the central normal metal is varied. However, an important difference between the two cases is that current carried under an applied bias between two normal metal reservoirs is a non-equilibrium current, whereas the Josephson current is an equilibrium current. In this paper, we study the Fabry-P\'erot interference exhibited by the equilibrium Josephson current.  

\section{Details of Calculation}
The Hamiltonian for a superconductor-normal metal superconductor junction is given by 
\begin{widetext}
 
\bea
H &=& \begin{cases} \Psi^{\dag}(x)\Big[\Big(-\f{\hbar^2}{2m}\f{\Do^2}{\Do x^2}-\mu \Big)\tau_z  + \De (\cos{\phi}\tau_x+~\sin{\phi}\tau_y)\Big]\Psi(x),~~ {\rm ~for~} ~x<0 \\ ~~\\
 \Psi^{\dag}(x)\Big[\Big(-\f{\hbar^2}{2m}\f{\Do^2}{\Do x^2}-\mu_0 \Big)\tau_z \Big]\Psi(x), ~~ {\rm for} ~~ 0<x<L ,  \\ ~~\\ 
 \Psi^{\dag}(x)\Big[\Big(-\f{\hbar^2}{2m}\f{\Do^2}{\Do x^2}-\mu \Big)\tau_z + \De \tau_x \Big]\Psi(x),  ~~{\rm for}~~  x>0, 
\end{cases} 
\label{eq:ham}
\eea
\end{widetext}

where $\Psi(x)=[c_{\ua}(x), ~c^{\dag}_{\da}(x), ~-c_{\da}(x), ~c^{\dag}_{\ua}(x)]^{T}$, and $c_{\si}(x)$ is annihilation operator for an electron of spin-$\si$ at $x$ and $\tau_x$, $\tau_y$, $\tau_z$ are Pauli spin matrices that act on the particle-hole sector. Here, $m$ is the effective mass of electrons, $\mu$ is the chemical potential in the superconductors, $\mu_0$ is the chemical potential in the normal metal region and can be tuned by an applied gate voltage, $\De$ is the strength of superconducting pair potential, and $\phi$ is superconducting phase difference. The wavefunction for this Hamiltonian is a four-spinor having the form $\psi=[\psi_{e,\ua}, \psi_{h,\da}, \psi_{e,\da}, \psi_{h,\ua}]^T$, where each of $\psi_{p,\si}$ ($p=e,h$, $\si=\ua, \da$) is a function of $x$ and $\psi_{p,\si}$ corresponds to an electron excitation of spin $\si$ for $p=e$, a hole excitation of spin $\si$ for $p=h$. 

The Hamiltonian does not have any spin dependent term. So, it has spin degenerate states. From eq.~\eqref{eq:ham}, it is evident that spin-up electrons mix with only spin-down holes and spin-down electrons mix with spin-up holes. So, the two-spinor eigenfunctions $\psi'=[\psi_{e,\ua}, \psi_{h,\da}]$ can be found, and the Josephson current determined from this can be multiplied with a factor of $2$ to obtain the total Josephson current.

The wavefunction $\psi'$ satisfies a probability current conserving boundary condition~\cite{soori20}. We choose the boundary condition that corresponds to a delta function impurity present at the junction between normal metal and the superconductor~\cite{btk}:
\bea  
\psi'(x_0^-) &=& \psi'(x_0^+), ~~~
\Do_x\psi'|^{x=x_0^+}_{x=x_0^-} = q_0 \psi'(x_0)  \label{eq:bc}
\eea 

The dispersion relation in the superconductors is $E=\pm\sqrt{(\hbar^2k^2/2m-\mu)^2+\De^2}$. Bound states appear at energies within the superconducting gap: $|E|<\De$. From the dispersion, $k$ can be found at energy $E$ and there are four complex solutions to $k$, of which two (say $k_1, k_2$) have positive imaginary part.  In the superconducting region  $x>L$, $k=k_1, k_2$ are taken and in the region  $x<0$, $k=-k_1, -k_2$ are taken as solutions so that the wavefunction is normalizable. The eigenfunction has the form: 
\bea  
\psi'(x) =\begin{cases} A_{1} e^{-ik_1x} \begin{bmatrix} u_{L,k_1} \\ v_{L,k_1}  \end{bmatrix} + A_{2} e^{-ik_2x} \begin{bmatrix} u_{L,k_2} \\ v_{L,k_2}  \end{bmatrix}, {~\rm for ~} x\le 0,  \\ 
  \begin{bmatrix} B_{e,f} e^{ik_ex} +B_{e,b} e^{-ik_ex} \\   B_{h,f} e^{-ik_hx} +B_{h,b} e^{ik_hx}   \end{bmatrix}, ~~~{\rm for ~~} 0\le x\le L,  \\ 
 C_{1} e^{ik_1x} \begin{bmatrix} u_{R,k_1} \\ v_{R,k_1}  \end{bmatrix} + C_{2} e^{ik_2x} \begin{bmatrix} u_{R,k_2} \\ v_{R,k_2}  \end{bmatrix}, {~~\rm for ~} x\ge L,  \\ 
\end{cases}
&& \label{eq:psi}
\eea
where $A_1, A_2, B_{e,f}, B_{e,b}, B_{h,f}, B_{h,b}, C_1, C_2$ are the coefficients that need to be determined, and  $[u_{P,k}, v_{P,k}]^T$ for $P=L, R$ and $k= k_1, k_2$ are the eigenspinors given by 
\bea  
\begin{bmatrix} u_{L,k} \\ v_{L,k}\end{bmatrix} &=& \begin{bmatrix} \De e^{i\phi} \\ E-\hbar^2k^2/2m +\mu \end{bmatrix}, \nn \\ 
\begin{bmatrix} u_{R,k} \\ v_{R,k}\end{bmatrix} &=& \begin{bmatrix} \De  \\ E-\hbar^2k^2/2m +\mu \end{bmatrix}.
\label{eq:spinors}
\eea

Using the boundary conditions in eq.~\eqref{eq:bc}, it can be shown that $M X=0$, where $X=[A_1, A_2, B_{e,f}, B_{e,b}, B_{h,f}, B_{h,b}, C_1, C_2]^T$ and $M$ is given by 
\begin{widetext}
\bea 
M &=& \begin{bmatrix}
       u_{L,k_1} & u_{L,k_2} & -1 & -1 & 0 & 0 & 0 & 0 \\
       v_{L,k_1} & v_{L,k_2} & 0 & 0 & -1 & -1 & 0 & 0 \\
       (i k_1-q_0)u_{L,k_1} & (i k_2-q_0)u_{L,k_2} & ik_e & -ik_e & 0 & 0 & 0 & 0 \\ 
       (i k_1-q_0)v_{L,k_1} & (i k_2-q_0)v_{L,k_2} & 0 & 0 & -ik_h & ik_h & 0 & 0  \\ 
       0 & 0 & e^{ik_eL} & e^{-ik_eL} & 0 & 0 & -e^{ik_1L}u_{R,k_1} & -e^{ik_2L}u_{R,k_2} \\ 
       0 & 0 & 0 & 0& e^{-ik_hL} & e^{ik_hL} &  -e^{ik_1L}v_{R,k_1} & -e^{ik_2L}v_{R,k_2} \\ 
       0 & 0 & -ik_ee^{ik_eL} & ik_ee^{-ik_eL} & 0 & 0 & (ik_1-q_0)e^{ik_1L}u_{R,k_1} & (ik_2-q_0)e^{ik_2L}u_{R,k_2} \\
       0 & 0 & 0 & 0 & ik_he^{-ik_hL} & -ik_he^{ik_hL} & (ik_1-q_0)e^{ik_1L}v_{R,k_1} & (ik_2-q_0)e^{ik_2L}v_{R,k_2} \\
      \end{bmatrix} \nn \\ 
      && 
\eea
\end{widetext}
This implies that the determinant of the matrix $M$ must be zero. We scan the energy range $-\De<E<\De$ and look for energies where ${\rm det.}(M)=0$. At such energies, one of the eigenvalues of $M$ turns out to be zero and the eigenvector $X$ corresponding to zero eigenvalue determines the eigenstate. The corresponding eigenstate given by eq.~\eqref{eq:psi} is numerically normalized. Now, the current carried by such a state can be found by calculating the current in the normal metal region~\cite{furusaki99}: $J'=(e\hbar/m){\rm Im}(\psi'^{\dag}\Do_x\psi')$. The current carried by both spin sectors is $J=2(e\hbar/m){\rm Im}(\psi'^{\dag}\Do_x\psi')$. The bound state energies come in pairs $\pm E_b$. There can be multiple pairs of energies at which  ${\rm det.}(M)=0$. The sum of currents carried by all the negative energy bound states gives the total Josephson current.
The Josephson current can also be calculated by the formula $J=4(e/\hbar)\sum_j\partial E_{b,j}/\partial\phi$, where $\pm E_{b,j}$'s are different bound state energies. We numerically find that both the methods give exactly the same Josephson current.

\begin{figure}[htb]
 \includegraphics[width=8cm]{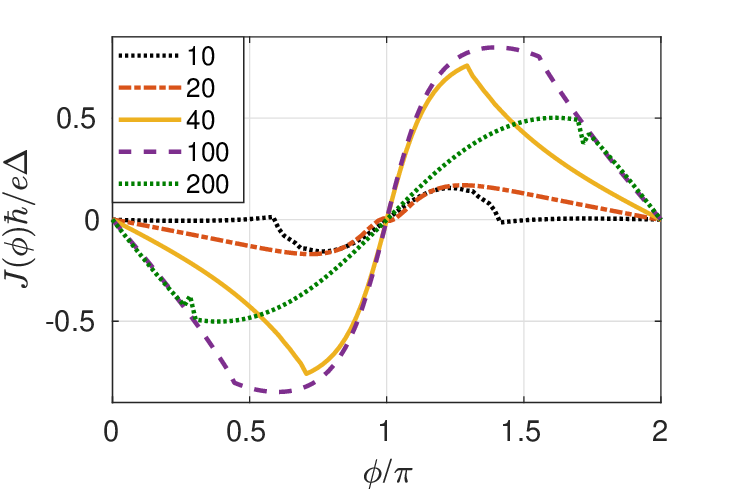}
 \caption{Current phase relation of a superconductor-normal metal-superconductor junction with $\mu=20\De$, $q_0=2\sqrt{m\De}/\hbar$, $L=10\hbar/\sqrt{m\De}$.  $\mu_0/\De$ is shown in the legend. }\label{fig:cpr}
\end{figure}

\begin{figure*}[htb]
 \includegraphics[width=12cm]{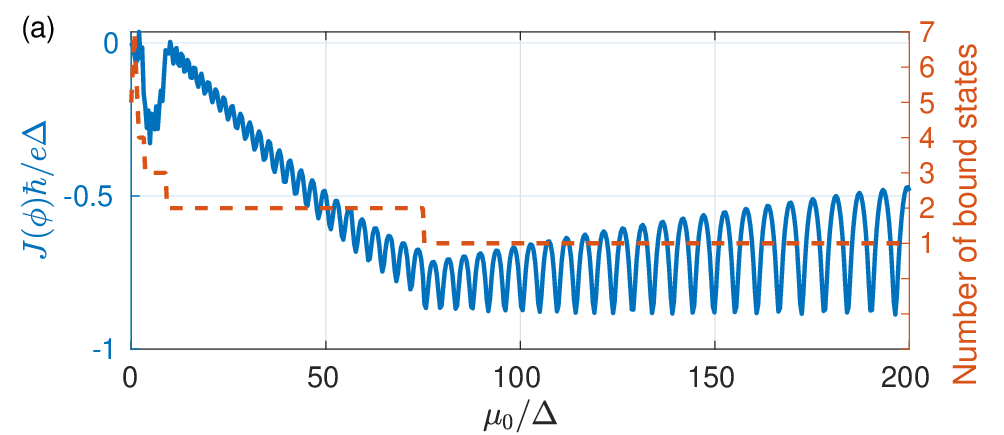}
 \includegraphics[width=5cm]{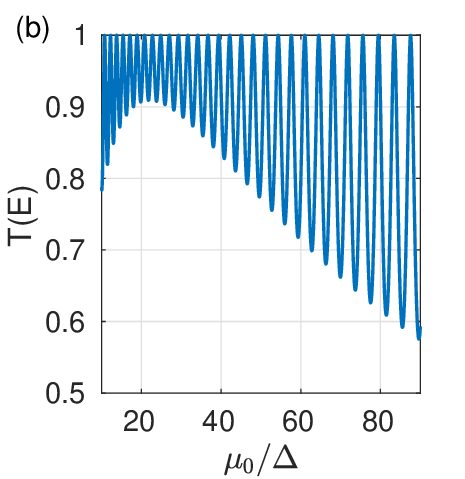}
 \caption{(a) Josephson current (on left y-axis) number of bound states (on right y-axis) versus $\mu_0$ for $\phi=\pi/2$.   (b) Transmission probability at zero energy versus $\mu_0$ for NM-NM-NM junction described by eq.~\eqref{eq:ham} in absence of terms proportional to $\De$. Parameters: $\mu=20\De$, $q_0=2\sqrt{m\De}/\hbar$, $L=10\hbar/\sqrt{m\De}$. }\label{fig:Jmu0}
\end{figure*}

\section{Results and Analysis}

We choose $\mu=20\De$, $q_0=2\sqrt{m\De}/\hbar$, $L=10\hbar/\sqrt{m\De}$ and numerically calculate the Josephson current $J$ as a function of the superconducting phase difference $\phi$ for different choices of $\mu_0$. The graph of current phase relation is shown in Fig.~\ref{fig:cpr}. The current is not a sinusoidal function of $\phi$. Interestingly, the current first increases as $\mu_0$ increases and then decreases. This motivates us to look at the dependence of Josephson current on $\mu_0$ at a fixed $\phi$. In Fig.~\ref{fig:Jmu0}, we plot the Josephson current versus $\mu_0$ for the same set of parameters. $\mu_0$ - the chemical potential in the normal metal region can be controlled in an experiment using an external gate voltage.  We find that the Josephson current is close to zero near $\mu_0=0$. This is because, the band bottom of the normal metal region is close to zero and at nonzero energies within the superconducting gap, for an electron plane wave state, there is no hole state which is plane wave in the normal metal region. As $\mu_0$ is increased further, the Josephson current increases in magnitude, but with oscillations. These oscillations are due to Fabry-P\'erot interference. If $\mu_{0,i}$ is the position of $i$-th local peak, $k_{e,i}\simeq \sqrt{2m\mu_{0,i}}/\hbar$ satisfies the relation $[k_{e,i+1}-k_{e,i}]L=\pi$ very well for $\mu_0>10\De$, since $|E|<\De$ can be neglected in comparison to $\mu_0$ in the expression $k_e=\sqrt{2m(\mu_0+E)}/\hbar$. This condition is the Fabry-P\'erot interference condition. 

Bohr-Sommerfeld like quantization refers to the fact that standing waves on a ring can have discrete momenta such that $\int p dx = nh$. But, in our case, the standing waves are not on a ring and the momenta $\pm\hbar k$ contribute to making the standing wave in the normal metal region. Hence, $\int p dx = \int_0^L\hbar k dx+\int_L^0(-\hbar k) dx = 2\hbar k L$ and the quantization condition implies $2\hbar k L =nh \implies (k_{n+1}-k_n)L=\pi$ which is the same as the Fabry-Perot interference condition.

Another feature of  Fig.~\ref{fig:Jmu0}(a), is that  around $\mu_0=70\De$, the maximum value of the magnitude of Josephson current at the peak saturates. This is because the number of bound states changes from $2$ to $1$ around $\mu_0=70\De$. The Josephson current is driven by the superconducting phase bias. 
% At the superconductor-normal metal interface, there is backscattering due to mismatch of the Hamiltonians on the two sides. When  two bound states carry the supercurrent, both the bound states experience the backscattering due to the interface. On the other hand,  when only one bound state carries the supercurrent, only one state experiences the resistance at the interface. 
We find that the supercurrent is higher in magnitude  when carried by only one bound state compared to when it is carried by two bound states. Also, the cusps in Fig.~\ref{fig:cpr} are due to change in the number of bound states at the location of the cusps when the phase difference is varied. 

Beyond $\mu_0\simeq 70\De$, the amplitude of oscillations increases as $\mu_0$ increases. To understand this feature, we look at the transmission probability in a similar NM-NM-NM junction as a function of $\mu_0$. NM-NM-NM junction can be described by the Hamiltonian in eq.~\eqref{eq:ham} by eliminating terms proportional to  $\De$. In Fig.~\ref{fig:Jmu0}(b), transmission probability at zero energy is plotted versus $\mu_0$ keeping other parameters the same. The transmission probability reaches $1$ at the peak, but the values at the local minima decrease as $\mu_0$ increases. This explains why the Josephson current shows oscillations with larger amplitude as the chemical potential $\mu_0$ increases. If $L$ is increased, the oscillations become more closely spaced, as can be understood from the FPI condition.

\section{Discussion}

These results hold when the length of the NM region is less than coherence length and the transport is phase coherent. At the same time, for the interference to happen, the length of the normal metal should be larger than a critical value given by $\pi\hbar/\sqrt{2m\mu_0}$. In typical superconductors, $\mu\gg \De$ and hence, we have chosen $\mu=20\De$. The barrier strength $q_0$ is assumed to be small compared to the Fermi wavelength so that the Josephson current is large. This holds for a smooth junction between normal metal and the superconductor. 

Instead of the dependence of the Josephson current on the chemical potential of the normal metal region, the dependence on the size of the normal metal region can be studied to probe the FPI in Josephson junctions. Such an experimental study supported by theoretical calculations using Gorkov formalism was performed by Gudkov et~al~\cite{gudkov}. Our analysis uses Bogoliubov de-Gennes formalism~\cite{furusaki99}. 

\section{Summary}
We have studied DC Josephson effect in a superconductor-normal metal-superconductor junction. We have written down the form of wavefunctions analytically and solved for the bound state energy, and coefficients numerically. We find current-phase relations and study the dependence of Josephson current on the chemical potential $\mu_0$ in the NM region. The oscillations in the Josephson current versus $\mu_0$ match with the Fabry-P\'erot interference condition. We have studied Fabry-P\'erot interference in equilibrium transport, in contrast to the Fabry-P\'erot interference commonly studied in nonequilibrium transport.  The number of bound states changes as $\mu_0$ is varied, and the current carried by one bound state can be higher in magnitude compared to the current by two bound states.
%We explain this feature with an argument rooted in backscattering at the interface.
Our results can be tested experimentally with present day technology.

\acknowledgements 
 AS thanks DST-INSPIRE Faculty Award (Faculty Reg. No.~:~IFA17-PH190),  SERB Core Research grant (CRG/2022/004311) and  University of Hyderabad Institute of Eminence PDF for financial support. 
\bibliography{ref_jjfp}

\end{document}